\documentstyle[preprint,aps]{revtex}
\draft
\tighten
\begin{document}
\title{Superstring  dominated  early universe and epoch dependent
gauge coupling.}
\author{\bf A. K. Chaudhuri\cite{byline}}
\address{ Variable Energy Cyclotron Centre\\
1/AF,Bidhan Nagar, Calcutta - 700 064\\}
\date{\today}
\maketitle
\begin{abstract}
We  have explored the possibility that the universe at very early
stage was dominated by (macroscopic) heterotic strings.  We  have
found  that  the dimensionless parameter $G\mu$ for the heterotic
strings varies from $10^{-2}$ to $10^{-4}$ as the universe evolve
from the matter dominance to radiation dominance. This led to the
interesting  consequence  of  epoch  dependent   gauge   coupling
constant. The gauge coupling constant at early times was found to
be   much   stronger   than   the   present  strong  interaction.
\end{abstract}

PACS numbers(s): 98.80.Cq, 11.17+y, 12.10.Gq\\

\section{Introduction}

Presently,   superstring   theory   \cite{sc82}  with  space-time
symmetry is being investigated by many authors. It is believed to
unify the gravitational force with the other forces of nature. It
can address some major unsolved problems in particle physics e.g.
origin of the spectrum  of  quarks  and  leptons,  the  hierarchy
problem,  i.e.  the  existence  of very small and very large mass
scales  etc.  Several  authors   have   explored   the   possible
cosmological    consequence    of    (macroscopic)   superstrings
\cite{wi85,tu88,mi92,ch94}. It is also conjectured that  bulk  of
the  dark  matter  may  consists  of  the lightest supersymmetric
particle \cite{go83}. Indeed, if superstring is  the  'theory  of
every  thing'  then  whole universe should be contained as one of
its solution. The most promising string theory is  the  heterotic
string  theory  \cite{gr85}.  The  theory  has  gauge  bosons and
fermions interacting through  a  large  unification  gauge  group
($E_8\times  E_8$). The theory even has natural handedness, which
translate into chirality of electroweak interaction. However, the
theory is Lorenze invariant only in ten dimensions, the extra six
dimensions of  which  can  be  compactified  on  some  Calabi-Yau
manifold.

In the cosmological context the dimensionless parameter $G\mu$ (G
is  the  gravitational  constant and $\mu$ is the string tension)
plays an important  role.  For  heterotic  strings,  there  is  a
fundamental  relation  between  the gravitational constant G, the
string  tension  $\mu$,  and  the  gauge  coupling   constant   g
\cite{tu88,gr85}:

\begin{equation}
G\mu=g^2/32\pi^2
\label{1a}
\end{equation}

\noindent  giving  $G\mu  \sim  10^{-3}$.  This is three order of
magnitude larger than that of cosmic strings  \cite{ki76}.  Is  a
universe  with cosmic fundamental strings with $G\mu\sim 10^{-3}$
viable? Most stringent bounds on  any  cosmological  model  comes
from  the observed micro wave back ground anisotropy \cite{ka84},
the pulsar timings \cite{vi81} and the  nucleosynthesis  analysis
\cite{da85}.  Strings  with $G\mu \sim 10^{-3}$ if present today,
would cause unacceptable distortions in the microwave background.
However, in the heterotic string theories the fundamental strings
become attached to axion  domain  walls  at  the  QCD  scale  and
thereafter  rapidly  disappear \cite{wi85}. It can then evade the
bounds imposed by the micro wave  background  anisotropy  or  the
pulsar  timings as they concern periods much later than QCD phase
transition. The nucleosynthesis bound require  that  the  gravity
waves  be  less  than  $18\%$  of  total  density  at the time of
nucleosynthesis. However, major fraction of gravity  waves  comes
from  periods  well  before  the  nucleosynthesis  time.  Roughly
speaking  there  is  equal  contribution  from  each  logarithmic
intervals  of  time.  For  superstrings, though we can ignore the
expansion after 100 MeV or so (QCD scale),  as  the  strings  are
formed  at  Plank energies ($10^{19}$ $GeV$) there is a 16 decade
of expansion to be considered. However,  unlike  cosmic  strings,
cosmic  fundamental  strings can decay into hundreds of mass less
states including the graviton \cite{wi90}. If one assume that the
decay is equally probable in every channel,  then  gravity  waves
can  be  reduced  by  a factor of hundred and it is expected that
nucleosynthesis bound can be  met  \cite{mi92,ch94}.  Considering
all  these  aspects,  in  an  earlier  paper \cite{ch94}, we have
argued that early universe can  be  heterotic  string  dominated.
However,  we  now  find  that the analysis was incomplete. Deeper
investigations revealed some interesting results, which  will  be
reported in the present paper. The paper is organised as follows:
in  section  2,  we  shall  describe briefly the evolution of the
string universe, the results will  be  discussed  in  section  3.
Summary will be given in section 4.

\section{Evolution of string dominated universe}

We  assume  that  just  after  the  Plank  time, the universe was
dominated  by  super  strings.  We  also  assume  that  all   the
compactification  has been completed, by the time universe starts
evolving. In string  theory,  the  gravitational  interaction  is
effectively  described  by  not  only  the metric but the coupled
system of metric and dilaton  field.  String  theory  also  shows
duality  i.e.  strings in very small space behave the same way as
the strings in very large spaces. However presently we ignore the
explicit dilaton field for the simple reason that we do not  know
its   value.   Also,   as   our   universe   begins   after   the
compactification, its dynamics may be  neglected.  For  the  same
reason,  we  ignore  the  duality  and  use Einstein's gravity to
describe the evolution of the universe \cite{tu88,mi92,ch94}.

How  a  superstring universe will evolve? Here we take a cue from
the cosmic string study \cite{vi81a}.  A  cosmic  string  network
quickly evolve into a scaling solution. Since action for both the
cosmic  string  and the cosmic superstring are purely geometrical
Nambu action, amenable to classical description, it  is  possible
that  cosmic  fundamental string network also quickly evolve into
scaling era,  provided  other  things  are  similar  \cite{tu88}.
Unlike  in cosmic string, there is no infinitely long superstring
in heterotic string theory. All the  strings  consists  of  loops
only.  However,  if  the  loops are very large $>>$ horizon, they
behave essentially like a infinitely long string. Thus  we  start
our universe initially consists of long string. They can chop off
loops, which in turn decay by emitting radiation. And we look for
scaling solution.

We assume the standard hot cosmological model with spatially flat
Robertson-Walker  metric  with  scale  factor  a(t). Expansion is
governed by the Einstein equation,

\begin{equation}
\dot{a}/a  = 8\pi G\rho/3 \label{1}
\end{equation}

\noindent  here  $\rho  =  \rho_m  +  \rho_{rad}$.  $\rho_m$  and
$\rho_{rad}$  are  the  matter  and  radiation   energy   density
respectively. We further divide $\rho_m$ into two parts, the long
string  energy  density  ($\rho_L$)  and  the loop energy density
($\rho_l$ ). We also define a radiation fraction r such that,

\begin{equation}
r = \rho_{rad} /(\rho_L + \rho_l + \rho_{rad}) \label{2}
\end{equation}

In   the  scaling  era  Einstein  eq.  can  be  solved  to  yield
\cite{mi92,ch94},

\begin{equation}
a(t) \propto t^{2/(r+3)} \label{3}
\end{equation}

The  Hubble  parameter  (H),  the  Hubble  radius ($R_H$) and the
horizon ($d_H$ ) can be obtained as,

\begin{eqnarray}
H =&& \dot{a}/a = 2/((r+3)t)\\
R_H=&& (r+3)t/2\\
d_H =&& a(t)\int dt' a(t')^{-1} = (r+3)t/(r+1)
\end{eqnarray}

The  main idea of scaling solutions \cite{vi81} is that, there is
a single length scale in the problem, which we  take  to  be  the
Hubble  radius  ($R_H$).  All  the other scales are determined in
terms of $R_H$ . We define a length scale $\xi$  on  the  string;
$\rho_L  =  \mu/\xi^2$.  As  long as the reconnection is frequent
between the strings, which should be the case for $\xi << R_H$  ,
it  will keep the network random so that $\xi$ will be related by
a constant factor  of  order  unity  to  the  typical  radius  of
curvature on the string. The existence of scaling solution can be
argued  thus:  when  $\xi << R_H$ , the long strings rapidly chop
off loops and the long string  density  falls,  and  $\xi$  grows
faster  than  $R_H$.  If $\xi >> R_H$ , then chopping off becomes
infrequent, the string density rises and $\xi$ falls relative  to
$R_H$ . It must be mentioned that the above discussion assume the
intercommuting probability to be unity.

In  the one scale hypothesis we can obtain analytical expressions
for  the  long  string,  loops  and  radiation  energy  densities
\cite{mi92,ch94}. For completeness purpose we briefly discuss the
procedure.

The  time evolution of long string energy density is described by
the equation \cite{mi92,ch94,al89},

\begin{equation}
\frac{d\rho_L}{dt} = - 3H\rho_L + H/R_H\xi \rho_L - C\rho_L/\xi
\label{5}
\end{equation}

\noindent  the  1st and 3rd term on the r.h.s. represent the loss
in long string energy density due to  the  Hubble  expansion  and
chopping  off of loops. The 2nd term represent the gain in energy
density due to stretching of strings as a result of expansion. In
the scaling era, eq.\ref{5} can be integrated to obtain the  long
string energy density as \cite{mi92,ch94,al89},

\begin{equation}
\rho_L = \gamma^2_s \mu/R^2_H \label{7}
\end{equation}

\noindent with the scaling density,
\begin{equation}
\gamma_s = \frac{1}{2C}[ r + \sqrt{r^2+ 4C}] \label{8}
\end{equation}

The  loop  production  in  the  scaling  era  is described by the
equation \cite{mi92,ch94,al89},

\begin{equation}
\frac{d\rho(l)}{dt}=-3H\rho(l)+\frac{\mu}{\xi^4}
f_{off}(l/\xi)-\frac{\mu}{\xi^4} f_{rec}(l/\xi) \label{9}
\end{equation}

\noindent  where $\rho(l)dl$ is the energy density partitioned in
loops of length l and l+dl. The  1st  and  3rd  terms  on  r.h.s.
represents the decrease in energy density due to Hubble expansion
and  reconnection  of loops with long strings. The 2nd term gives
the gain in energy density due to chopping off of  long  strings.
The chopping efficiency C, satisfies the energy sum rule,

\begin{equation}
C = \int dx f(x) = \int dx (f_{off}(x)-f_{rec}(x)) \label{10}
\end{equation}

From  statistical  mechanical  considerations, Albrecht and Turok
\cite{al89}, obtained the chopping off function as,

\begin{equation}
f_{off}(x) = A x^{-1/2} exp(-Bx)    \label{11a}
\end{equation}

\noindent  with  A and B smoothly interpolating between radiation
and matter era.

The reconnection function can be written as \cite{al89},

\begin{equation}
f_{rec}(l/\xi) = k \xi^2 l \rho(l)/\mu \label{11b}
\end{equation}

\noindent  by  noting that the probability per unit time, for any
loop of length l to hit a long string is $\approx kl/\xi^2$ (with
long string density $\approx \xi^{-2}$ ). Here k is a  factor  of
order unity, determining the geometrical cross section for a loop
to hit the long string.

The  slow decay of loops into gravitational radiation can also be
taken into account. In analogy with  cosmic  string,  the  energy
loss  per unit time due to gravitational radiation can be assumed
to  be  $\Gamma   G\mu^2$   \cite{ki76}   ,   with   $\Gamma$   a
characteristic   constant   for  the  loops.  Eq.\ref{9}  can  be
integrated  to  obtain  the  energy  density  of  the  loops   as
\cite{ch94},

\begin{equation}
\rho_l      =A(r)\frac{2r}{r+3}\gamma_s^3      B^{\frac{2r}{r+3}}
\int^{x_2}_{x_1} dx  x^{-3(r+1)/(r+3)}I_0(\frac{R_Hx}{B\gamma_s})
\mu/R^2_H
\label{15}
\end{equation}

\noindent where,
\begin{equation}
I_0(\frac{R_Hx}{B\gamma_s})= \int^\infty_0 (z+\frac{x}{B})
exp[-( B + \frac{2k\gamma_s}{r+3})] dz     \label{16}
\end{equation}

and
\begin{eqnarray}
x_1= 2B\gamma_s/(r+3)\\
x_2= x_1(1+\frac{r+3}{(r+1)\Gamma_{eff}})\\
\Gamma_{eff}= \Gamma G\mu
\end{eqnarray}

Energy  density  of  radiation  emitted  from  the  loops  can be
calculated easily \cite{mi92,ch94,al89} and  we  write  down  the
result only. At a time $t >> t_{Plank}$ ,

\begin{equation}
\rho_{rad} =A(r) \frac{2}{(1-r)(r+3)} \gamma_s^4
B^{\frac{3r+3}{r+3}}\Gamma_{eff}X_0 Y_0 \mu/R^2_H \label{18}
\end{equation}

\noindent where,
\begin{eqnarray}
X_0=\int x^{\frac{-4r-6}{1+r}} exp(-x) I_0[R_Hx/B\gamma_s]\\
Y_0=\int  g(x)x^{\frac{1-3r}{1+r}}dx\\
g(x) = x^7 e^{-x}/6!
\end{eqnarray}

We  now  have  all  the  relations  needed to evaluate the energy
density of the universe. The undetermined parameters of the model
are, the chopping efficiency C,$\Gamma_{eff} (=\Gamma G\mu)$  and
k,  the geometric factor of loop reconnection function. They will
be determined in the next section.

\section{Results and discussions}

The  chopping  efficiency  (C)  of cosmic fundamental strings are
unknown. Even for cosmic strings, the chopping efficiency is  not
determined  well. Thus while the simulation study of Albrecht and
Turok \cite{al89} gave C=0.74,  much  larger  value,  C=0.16  was
obtained  in  the  simulation  studies  of  Bennett  and  Bouchet
\cite{be89}. In  our  previous  analysis  \cite{ch94},  both  the
values were used. However, it is possible to obtain the (maximum)
chopping  efficiency  for  cosmic  fundamental strings in the one
scale  hypothesis  \cite{ki76}.   The   essential   'one   scale'
assumption  is  that  the time scale for loops production is just
the scale length $\xi$ of the infinite strings \cite{ki76}  (i.e.
C  is  a  constant). This has been verified to a good accuracy in
the simulation studies also \cite{al89}. Then using the sum  rule
eq.(\ref{10})  and  demanding that C remain constant at all r, we
can find k. This induces r-dependence on  k.  We  further  demand
that  k,  being  a  geometric  parameter, should be positive. The
maximum value of C for which k remain positive at all r is  found
to be C=0.153. This is the maximum chopping efficiency for cosmic
fundamental  strings.  Chopping efficiency greater than this lead
to unacceptable, negative k. We note that  the  maximum  chopping
efficiency  thus  obtained is very close to the value obtained by
Bennet and Bouchet \cite{be89}, in their simulation studies.

As  mentioned  earlier, the geometric cross section k now depends
on the radiation fraction  (r).  In  fig.1,  we  have  shown  the
variation  of  k  with  the  radiation  fraction  for  C=0.153. k
increases with the radiation fraction. The variation of k with  r
can  be  understood  easily. As r increases, string density drops
and geometric cross section k has to increase  to  keep  chopping
efficiency fixed.

The  other parameter of the model ($\Gamma_{eff}=\Gamma G\mu$) is
obtained    from    the    self-consistency    condition     that
\cite{mi92,ch94},

\begin{equation}
r(\rho_L +\rho_l+\rho_{rad}) -\rho_{rad} = 0 \label{21}
\end{equation}

The  condition  demand  that  the  motion  of the strings and the
radiation in the back-ground space-time determines the space-time
itself \cite{mi92}. This aspect reflects the  fundamental  nature
of string theory that the space-time is unified with matter.

In  fig.2, we have shown the radiation fraction (r) as a function
of $\Gamma_{eff}$. As $\Gamma_{eff}$ increases, r increases, till
a maximum value $r_{max} \approx 0.8$ is  reached.  With  further
increase  of  $\Gamma_{eff}$,  r decreases. We thus find that for
two  values  of  $\Gamma_{eff}$,  the  universe  can  stay  at  a
particular  radiation  fraction.  One  can  then  distinguish two
regions, in region I (below $\Gamma_{eff} = 10^{-3}$),  radiation
fraction  increases  with  $\Gamma_{eff}$ and in region II (above
$\Gamma_{eff}=10^{-3}$),  it  decreases.  If  we  can  form   the
universe  at  $r_{max}\approx0.8$,  then  it  can  go either way.
Physics will be different in different  region.  Later  we  shall
argue  that the region I is unphysical. The finding that $r_{max}
\sim 0.8$, is in agreement with  ref.\cite{tu88}.  Radiation  can
not increase indefinitely in presence of strings.

In  fig.3,  we have shown the variation of long string, loops and
radiation energy density, with the radiation  fraction  (r).  The
two  regions  are  shown  separately.  In both the region, energy
densities of different  components  of  the  universe  behave  in
similar  fashion.  They  increase  with r. As expected the energy
density of radiations picks up very fast. However, in  region  I,
energy densities attained are much higher than in region II. Now,
there  is another self-consistency condition arising from the use
of the flat universe model. For the flat  universe,  the  density
should be equal to the critical density,

\begin{equation}
\rho   =  \rho_{crit}  =  \frac{3}{8\pi  G\mu}
\mu/R^2_H \label{22}
\end{equation}

This   relation  can  be  used  to  obtain  $\Gamma$  and  $G\mu$
separately. In fig.4, variation of $G\mu$ and $\Gamma$  with  the
radiation  fraction  are shown. In both the region, dimensionless
constant $G\mu$ decreases as the radiation fraction increases. As
$G\mu$ is related to the  coupling  constant  (see  eq.\ref{1a}),
this  indicate  that in a string dominated universe, the coupling
constant is epoch dependent. Before we elaborate on this  result,
let  us  note  the  behavior  of the decay constant $\Gamma$. The
behavior of $\Gamma$ in region I and II are completely different.
While in region I, $\Gamma$ increases exponentially by six orders
of magnitude as the radiation fraction increases from 0  to  0.8,
it  remains  approximately  constant  in region II. In fig.5, the
same result is shown in a different fashion. Here, we have  shown
the  decay  constant  $\Gamma$ as a function of the dimensionless
constant $G\mu$. In region I, $\Gamma$ varies from  $10^{-5}$  to
$10$  as  $G\mu$ increases from $10^{-4}$ to $10^{-2}$. In region
II, for the same range of $G\mu$, $\Gamma$ remains  approximately
constant.  Now  $\Gamma$  is  characteristic  of the loops and we
expect it to be constant and independent of $\mu$, as obtained in
region II. We thus argue that the region I is unphysical and only
region II is physical and need to be considered. That  in  region
II, $\Gamma$ remain constant gave credence to our model. Value of
$\Gamma  \approx$30,  obtained in the present calculation is also
interesting. This value is similar to the value obtained  in  the
cosmic  string  case  for  gravity  wave emission \cite{ki76}. As
cosmic fundamental strings can decay into  hundreds  of  massless
states, two order of magnitude larger $\Gamma$ is expected. Small
value of $\Gamma$ indicate that the zero mode emission for cosmic
fundamental  strings  is  not efficient. This may have bearing on
lightest supersymmetric particle as a candidate for dark matter.

The  epoch  dependent  gauge  coupling  obtained  in  the present
calculation can have several  interesting  consequences.  Let  us
first  see whether the result is physical or not? We note that we
do not have explicit dilaton  field  in  the  theory.  In  string
theory,  the  coupling  constant is determined by the expectation
value of the dilaton field. Apparently,  it  is  surprising  that
though  the  theory  do not have explicit dilaton field, it gives
epoch dependent gauge coupling. However, the  theory  do  contain
expectation value of the dilaton field through the string tension
$\mu$.  The epoch dependent coupling constant also have a natural
explanation. As the universe  expands,  the  strings  has  to  be
stretched    to    keep   themselves   cosmologically   relevant,
consequently $\mu$ decreases. In quantum field theory  also,  the
coupling  constants  are  effectively  constant  only  at certain
energy. They are energy  dependent.  In  an  expanding  universe,
temperature  of  the  universe  can be considered as the relevant
energy scale for the coupling  constant.  Then  as  the  universe
expands,   radiation   fraction  increases,  temperature  of  the
universe decreases, resulting  into  the  epoch  dependent  gauge
coupling constant.

We  note  that,  epoch  dependent  coupling  was  obtained in our
earlier analysis \cite{ch94} and also in the analysis of Minakata
and Mashino \cite{mi92}. It was not elaborated upon. Tsetylin and
Vafa \cite{ts92} also obtained  a  similar  result  in  a  recent
calculation.  They  have  considered string cosmology taking into
account of dilaton field and string duality. One of the result of
their paper is that the dilaton field is not a constant field but
evolve. They have not elaborated on this point,  but  the  result
indicate an epoch dependent coupling constant.

There  is  a close similarity of our result with the Large number
theory  (LNT)  of  Dirac.  In  LNT  also,  Dirac  found  that   a
combination  of  G  and  fundamental particle masses changes from
epoch to epoch, requiring either G or fundamental particle masses
to be epoch dependent. Dirac {\em choose} to  keep  the  particle
masses  fixed and allow the gravitational constant to change from
epoch to epoch. Interestingly, in the present model  also,  in  a
different  context,  we  find  that  $G\mu$ changes from epoch to
epoch. Since we are compelled to keep G fixed (Einstein  theory),
we  have  to  change  $\mu$,  leading  to  epoch  dependent gauge
coupling constant.

In  fig.6,  we  have  shown  the variation of the string coupling
constant as a function of the radiation  fraction  for  both  the
region  I  and  II.  The  coupling  constant  was  calculated as,
\cite{mi92},

\begin{equation}
\alpha=g^2/{4\pi}=8\pi G \mu
\end{equation}

Behavior  of  $\alpha$  in  I  and II are similar, only magnitude
differ. In the physical region II, as the  universe  evolve  from
matter to radiation dominance, the coupling constant changes from
a  high  value  of  $\sim$0.46 to a low value of $\sim$0.005. For
comparison purpose, we have shown the  three  coupling  constants
($\alpha_1,\alpha_2$   and  $\alpha_3$)  of  the  standard  model
\cite{lep}. Also shown is the currently accepted  standard  model
unified coupling constant ($\alpha_{GUT}=\approx 1/24$ at the GUT
scale.  Interactions  at  early times were very much strong, much
stronger  than  the  present  day  strong  interactions.  As  the
universe  expands,  the  interaction  become weaker. The standard
model unified interaction is  reached  at  $r  \approx  0.6$.  It
becomes  further  weakened  as universe expands and become weaker
than the (present) weak interaction at $r=r_{max}$ $\approx 0.8$.

Let us now consider some implications of an epoch dependent gauge
coupling  constant.  If  the  gauge  coupling is epoch dependent,
several interesting things can happen. For example,  spectrum  of
states of strings is given by \cite{al86},

\begin{equation}
w(m)=w_0(m\sqrt{\alpha\prime})^{-a} e^{bm\sqrt{\alpha\prime}}
\end{equation}

\noindent where $\alpha\prime$ is the Regge slope, related to the
string tension $\mu$ or the dimensionless parameter $G\mu$ as,

\begin{equation}
\alpha \prime = \frac{1}{2\pi \mu} = \frac{1}{2\pi G\mu m^2_{pl}}
\end{equation}

As  $G\mu$  changes  by  two  order  of magnitude as the universe
evolve from matter to radiation dominance, the density of  states
will  change.  However,  large  Plank  mass  will ensure that the
density of states of low mass particles  remain  unchanged,  only
that  of  heavy  particles  will  differ  from  epoch  to  epoch.
Particles  with  mass  $m\sim  m_{pl}$  will  be  more  copiously
produced   in   early  universe.  As  the  heavy  particles  will
ultimately decay into lighter ones, we can say that  the  density
of  states  of  all  the particles will be affected, if the gauge
coupling is  epoch  dependent.  Though  at  present,  it  is  not
possible  to  calculate  the  fundamental  particle masses in the
string theory, possibility can not be ruled out that  fundamental
particle  masses  are  also epoch dependent. This also raises the
possibility of verifying the hypothesis of string dominated early
universe. As the string universe has to  decay  away  before  QCD
transition  we  can  not hope to obtain experimental signature of
string domination of the universe. And as yet there seems  to  be
no other independent way of verifying the hypothesis. However, as
the  nucleosynthesis  analysis concerns periods much earlier than
the time of nucleosynthesis, if the fundamental  particle  masses
and  mass  spectra were different at early times, its results may
change. It is then possible that early string dominated  universe
will left its imprint on the nucleosynthesis analysis.

Though  as  we  have  argued above, it is possible that the epoch
dependent gauge coupling constant is physical  and  the  universe
was  string dominated, we must mention the other possibility. The
present model relies  on  the  assumption  of  scaling  solution.
However, even in cosmic strings we find that there is controversy
about  the scaling law. While the simulations studies of Albrecht
and Turok \cite{al89} agree with the one scale  hypothesis,  that
of  Bennet  and Bouchet \cite{be89} disagree. It is then possible
that the assumption of scaling solution  for  fundamental  string
network  is  wrong.  The  epoch  dependent  gauge coupling is the
manifestation of the wrong assumption and the early universe  was
not superstring dominated.

\section{Summary}

To summarise, we have studied the evolution of a string dominated
early  universe.  The evolution was followed in a flat space-time
within the one scale hypothesis. The long  string  can  chop  off
loops,  which  in  turn can decay by emitting radiation. They can
also get reconnected with the long string. The  reconnection  was
treated  in a manner consistent with the one scale hypothesis. In
the one scale model, chopping efficiency of  long  strings  is  a
constant.  This  condition  was  utilsed  to  obtain  the maximum
chopping efficiency, which is very close to the value obtained by
Bennet and Bouchet \cite{be89} in a simulation  study  of  cosmic
strings.  For this chopping efficiency, imposing self-consistency
condition that the motion of strings  determine  the  space-time,
and  that  the  density be equal to the critical density, we have
obtained the dimensionless constant $G\mu$ for  the  strings  and
its decay characteristics. We confirmed that radiation density is
limited  in  presence  of  strings,  a  result  obtained by Turok
\cite{tu88}. We also found that, as the  string  universe  evolve
from  matter  to  radiation  dominance,  dimension less parameter
$G\mu$ varies from $10^{-2}$ to  $10^{-4}$.  This  leads  to  the
interesting   consequence   of  epoch  dependent  gauge  coupling
constant. This is also in agreement with the finding of  Tseytlin
and  Vafa  \cite{ts92}  that  dilaton  field evolve. We also find
close similarity of the present result with Dirac's Large  number
theory.  Epoch  dependent  gauge  coupling can results into epoch
dependent fundamental particle masses.  Density  of  states  will
also  be  epoch  dependent.  It  also raises the possibility that
effect  of  string  dominated  universe  will  be  felt  on   the
nucleosynthesis analysis.

\begin{figure}
\caption{   Variation   of   the  geometric  factor  k  for  loop
reconnection with long strings, with radiation fraction (r)  for
chopping effeciency C=0.153.}
\end{figure}

\begin{figure}
\caption{Radiation     fraction     (r)    as   a   function   of
$\Gamma_{eff}$ as obtained
from the self-consistent condition.}
\end{figure}

\begin{figure}
\caption{The  long  string  ($\rho_L$ ), loop ($\rho_l$ ) and the
radiation ($\rho_{rad}$ ) energy density, in units of $\mu/R^2_H$
, as a function of radiation fraction (r)}
\end{figure}

\begin{figure}
\caption{The variation of $G\mu$ and $\Gamma$ with radiation fraction
obtained from the self-consistent condition $\rho=\rho_{crit}$.}
\end{figure}

\begin{figure}
\caption{  The relation between $\Gamma$ and $G\mu$ obtained from
the   self-consistent   condition   $\rho   =   \rho_{crit}$   .}
\end{figure}

\begin{figure}
\caption{   The  string  coupling  constant  $\alpha$  as  a
function  of  the  radiation  fraction  (r). The straight lines
are  drawn  to  show  the  strength  of  the three standard model
coupling constants at $M_z$ and their unified  value  at  the  GUT
scale.}
\end{figure}
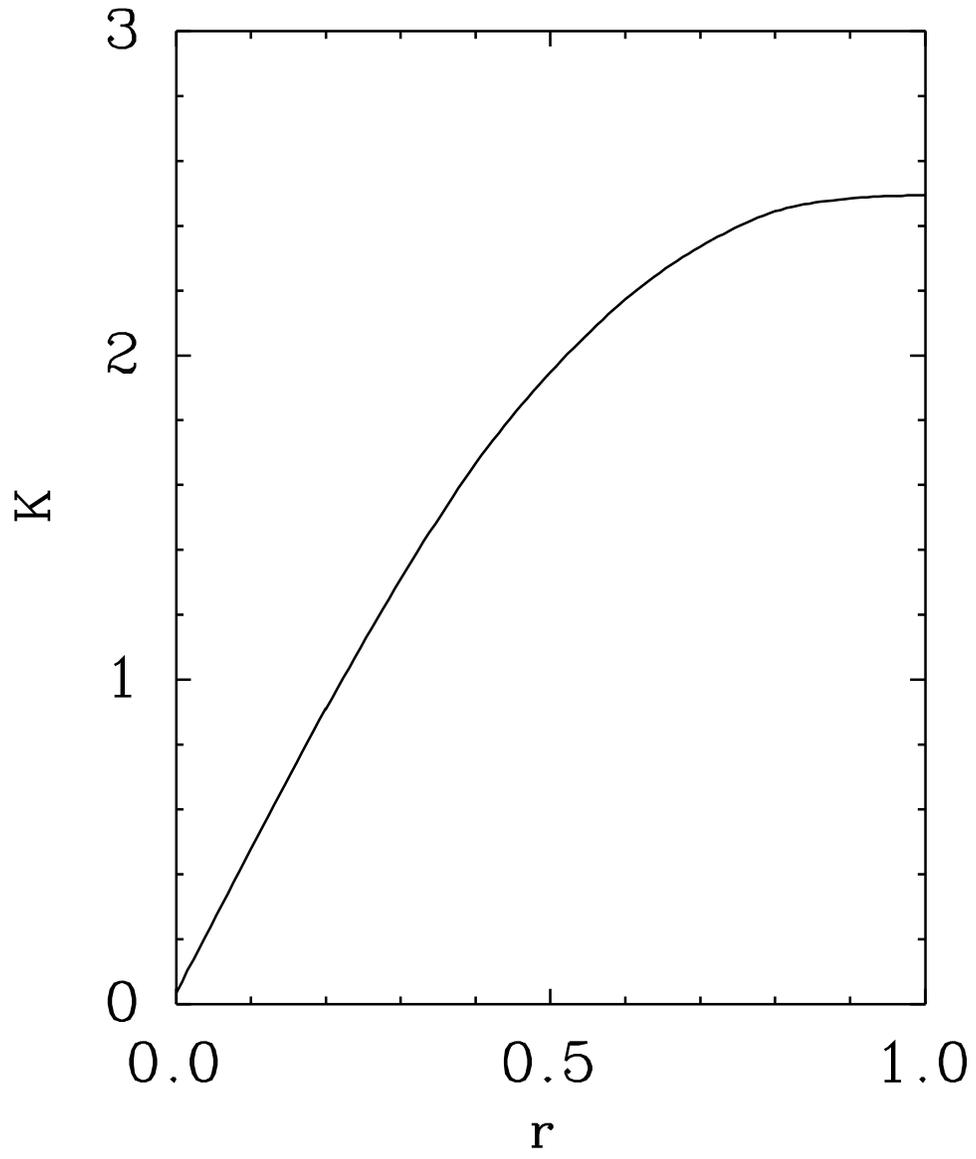
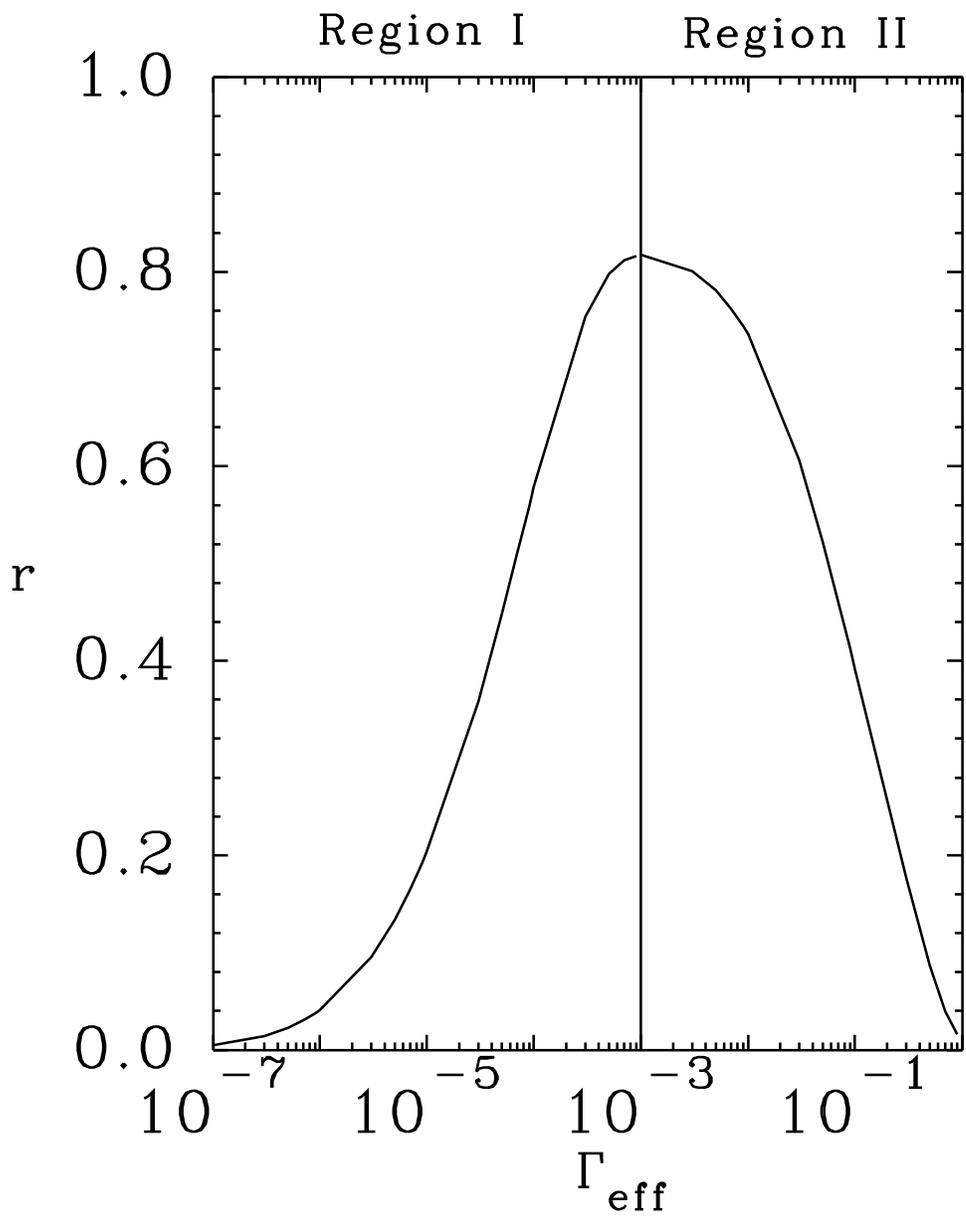
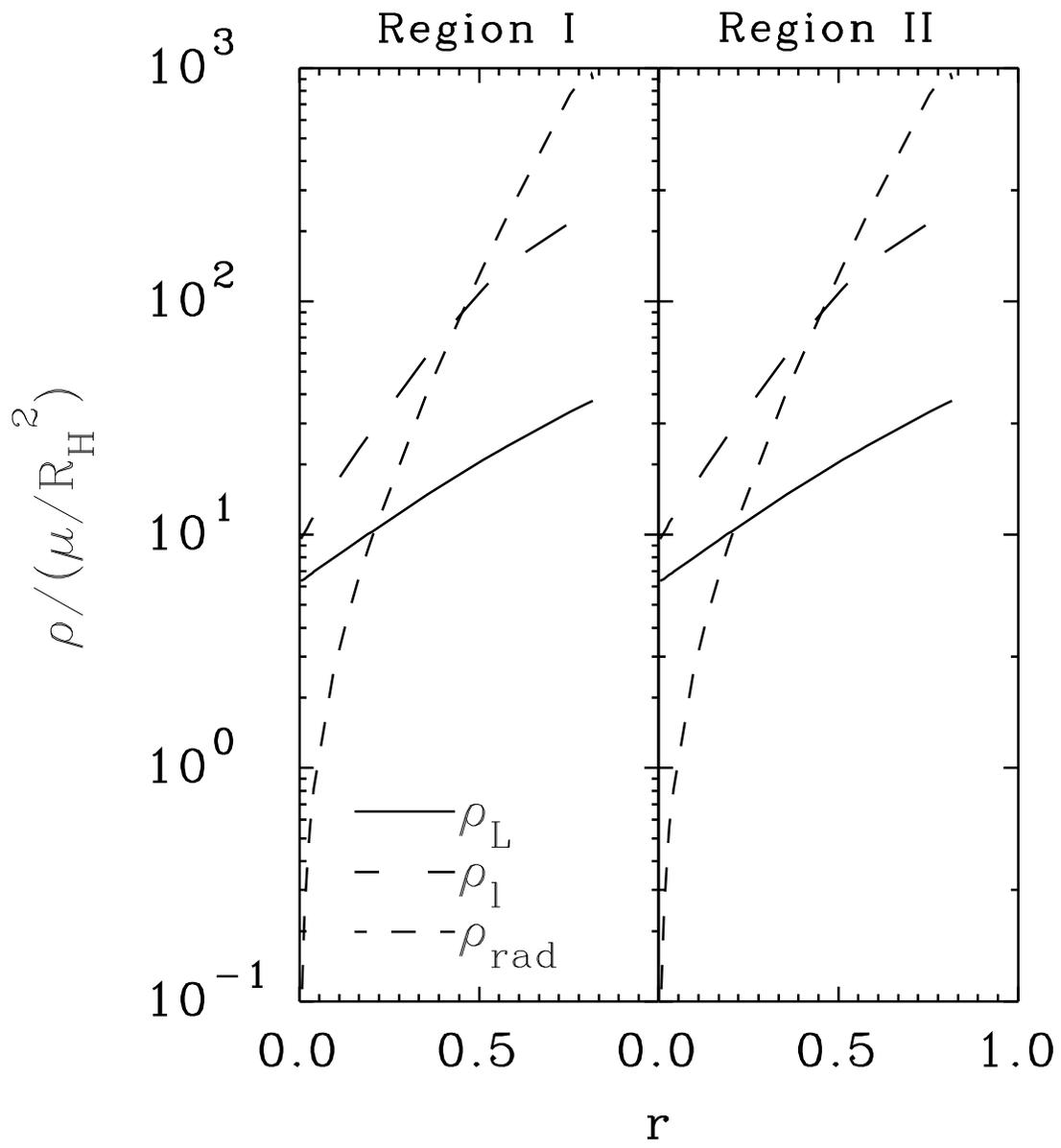
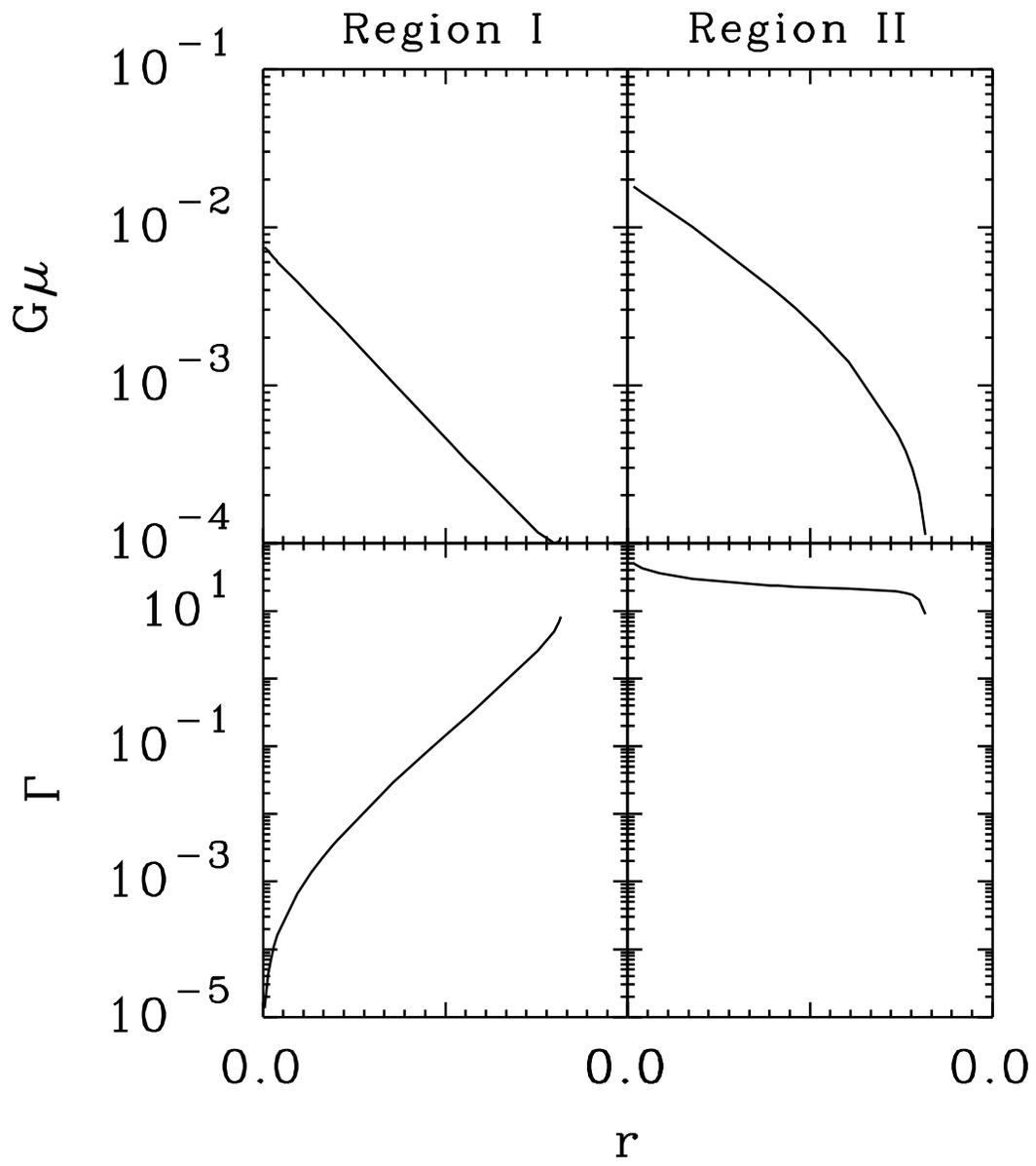
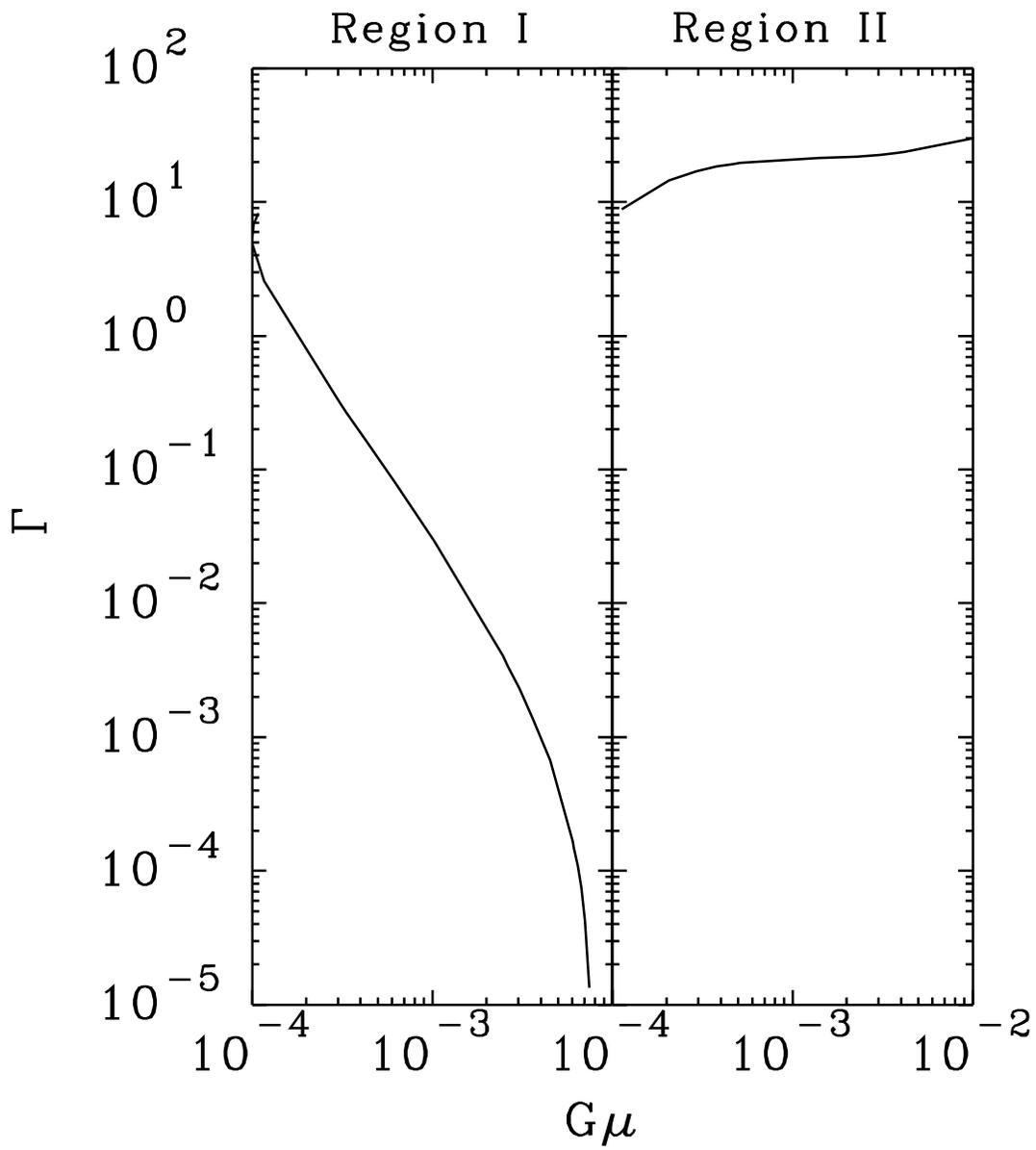
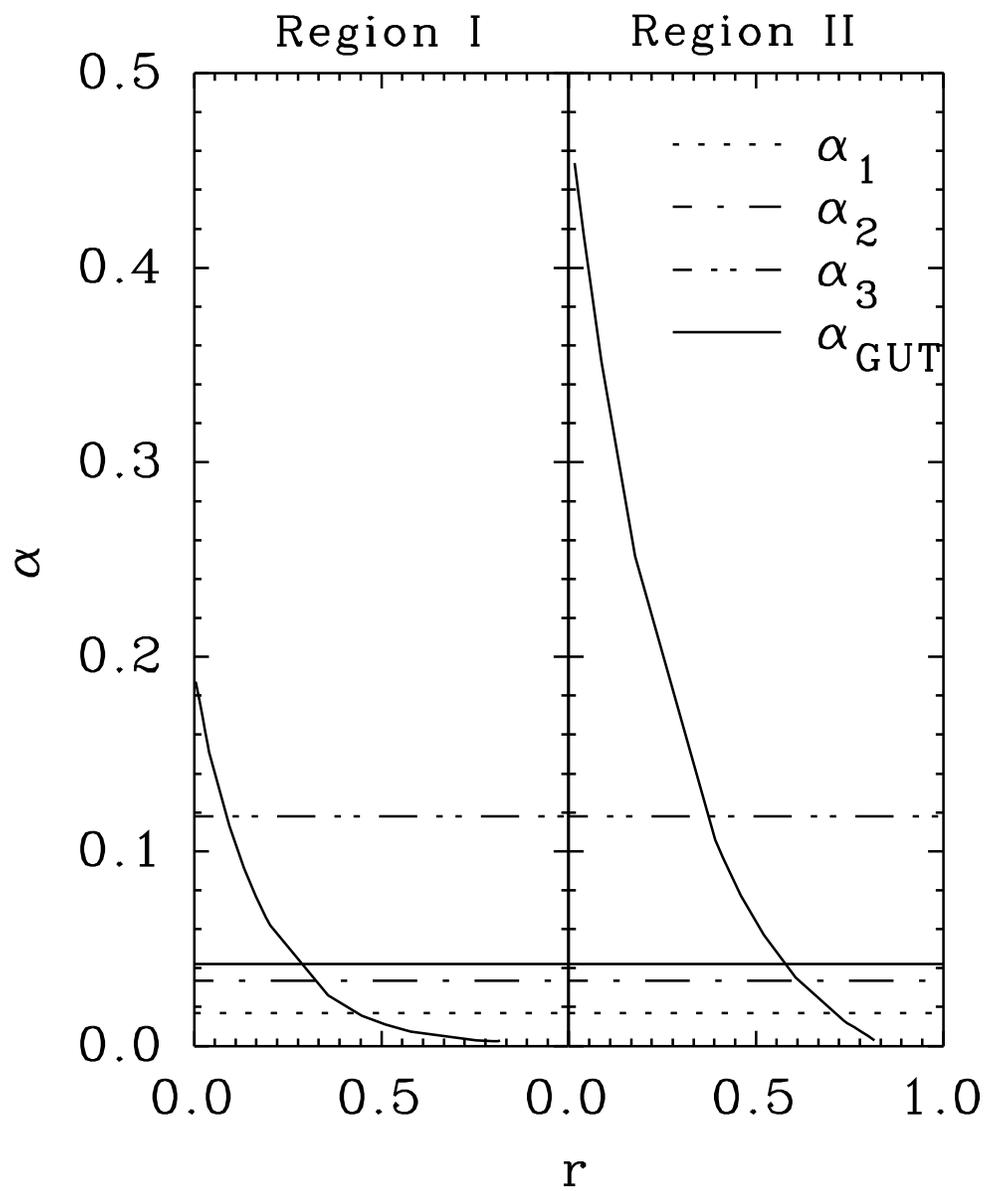
\end{document}